\documentclass[12pt, twoside]{article}

\usepackage{amsfonts}
\newcommand{\bb}{\begin {minipage} {3cm}\begin{center}}
\newcommand{\ee}{\end{center}\end{minipage}}
\newcommand{\bc}{\begin {minipage} {2.5cm}\begin{center}}
\newcommand{\bd}{\end{center}\end{minipage}}

\newcommand {\C} {{\mathbb C}}
\newcommand {\R} {{\mathbb R}}

\newcommand {\q}{\begin{quote} \small}

\newcommand {\be}{\begin{equation}}
\newcommand {\e}{\end{equation}}
\newcommand {\bea}{\begin{eqnarray}}
\newcommand {\ea}{\end{eqnarray}}
\newcommand {\g}{{\mathfrak g}}
\newcommand {\ga}{\gamma}
\newcommand {\fract}[2]{\mbox{${\textstyle{\frac{#1}{#2}}}$}}
\newcommand {\latop}[2]{#1\atop #2}
\newcommand {\tatop}[2]{\mbox{${\textstyle{\latop{#1}{#2}}}$}}
\begin{document}
\newtheorem {lemma}{Lemma}[subsection]
\newtheorem {theorem}{Theorem}[subsection]
\newtheorem {coro}{Corollary}[subsection]
\newtheorem {defi}{Definition}[subsection]
\newtheorem {obs}{Remark}[subsection]
\newtheorem {prop}{Proposition}[subsection]
\newtheorem {exa} {Example} [subsection]

\begin{flushright}
FTUV-00-0308 , IFIC-05-00\\ 
DAMTP-2000-17\\
\vskip .5cm
March 8, 2000
\end{flushright}
\vskip 2cm

\begin{center}
{\Large Fermionic realisations of simple Lie algebras} \\
\vspace{0.1cm}
{\Large and their invariant fermionic operators}\\
\vspace{1cm}

\begin{sl}
{\large J.A. de Azc\'arraga$^1$ and A.J. Macfarlane$^2$}\\
\vskip .5cm
$^1$Dpto. de F\'{\i}sica Te\'orica and IFIC, Facultad de Ciencias,\\
46100-Burjassot (Valencia), Spain\\
$^2$Centre for Mathematical Sciences, D.A.M.T.P\\
Wilberforce Road, Cambridge CB3 0WA, UK \\
\end{sl}
\end{center}
\vspace{1cm}
\begin{abstract}
We study the representation ${\cal D}$ of a simple compact Lie
algebra $\g$ of rank $l$ constructed with the aid of 
the hermitian Dirac matrices of a (${\rm dim}\,\g$)-dimensional
euclidean space. The irreducible representations of $\g$ contained 
in ${\cal D}$ are found by providing a general construction on suitable 
fermionic Fock spaces. We give full details not only for the 
simplest odd and even cases, namely $su(2)$ and $su(3)$, but also 
for the next ($\mbox{dim}\,\g$)-even case of $su(5)$. 
Our results are far reaching: they apply to
any $\g$-invariant quantum mechanical system containing
${\rm dim}\,\g$ fermions. Another reason for undertaking this study 
is to examine the role of the $\g$-invariant fermionic operators 
that naturally arise. These are given in terms of products of 
an odd number of gamma matrices, and include, besides a
cubic operator, ($l-1$) fermionic scalars
of higher order. The latter are constructed from the Lie algebra
cohomology cocycles, and must be considered to be of theoretical
significance similar to the cubic operator. In the 
(${\rm dim}\,\g$)-even case, the product of all
$l$ operators turns out to be the chirality operator
$\gamma_q,\; q=({{\rm dim}\,\g+1})$.

\end{abstract}

\section{Introduction}

   Let $G$ be a simple compact Lie group, of dimension $r$ and rank $l$,
of Lie algebra $\g$ defined by
\be \label{A1}
      [X_a \, , \, X_b]=iC_{abc} \, X_c  \quad ,
\e \noindent
where $a,b,c = 1,2, \dots , r$, and the 
generators $X_a$ are hermitian. We wish to examine here the 
realisation $X_a \mapsto S_a$ of $\g$ given by ~\cite{brra} 
\be \label{A2}
S_a= -\fract{1}{4} \, i C_{abc} \ga_b \ga_c \quad ,
\e \noindent
where the $\ga_a$ are the hermitian Dirac matrices

\be
\label{A3} \{ \ga_a \, , \, \ga_b \} =2\delta_{ab} 
\e 

\noindent
of a euclidean space of dimension $r\equiv\mbox{dim}\,\g$. 
We wish to describe 
exactly which representation ${\cal D}$ of $\g$ is provided 
by (\ref{A2}), and how the vector space ${\cal V}_{\cal D}$ 
which carries it may be constructed explicitly. We shall 
address first (and more fully) the even case in which both
$r$ (and hence the rank $l$) are even, contrasting it afterwards 
with the somewhat different, but also allowed and interesting,
odd case in which $r$ (and $l$) are odd.
 
We have a general reason for undertaking this study and one
specific class of applications immediately in mind.
The reason is the following: the Fock space of any 
quantum mechanical system in which there exists $\mbox{dim}\,\g$ 
fermions transforming according to $G$ is, necessarily, the 
carrier space of the representation ${\cal D}$ under 
investigation here. The specific
application that provides much of our motivation
is the study of the hidden supercharges $Q_s$ in $G$-invariant
supersymmetric quantum mechanical systems involving 
a set of exactly $\mbox{dim}\,\g$ fermions. There is one such 
$Q_s$ for each Lie algebra cohomology cocycle of $\g$
and its construction involves the associated completely 
antisymmetric invariant tensors fully contracted with the 
fermionic variables. These matters will be treated
in a forthcoming publication.

The first result of this paper
(Sec. 2) is a rather intriguing one: in the {\it even} case,
${\cal D}$ is the direct sum of $2^{\fract{l}{2}}$ 
copies of the irreducible representation (irrep) of $\g$ 
whose highest weight $\Lambda$ is equal to the Weyl 
principal vector $\delta =(1,\mathop{\dots}\limits^l,1)$. 
Here the Weyl principal vector of $\g$, equal to half 
the sum of the $(r-l)/2$ positive roots of $\g$, has been 
referred to a basis of the fundamental dominant weights of $\g$. 
Below, in Sec. 2, we establish this result, giving a 
systematic description of ${\cal V}_{\cal D}$ as a 
{\it fermionic} Fock space. 

Sec. 3 gives
full treatment of the $\g=A_2=su(3)$ case, 
which clarifies many of the issues typical of the even case. 
In particular, we exhibit the role of both the chirality 
operator and of the Kostant fermionic $SU(3)$-invariant 
cubic operator $K_3$. This operator is defined by
\be
\label{A4} K_3 =-\fract{1}{12} i C_{abc} \ga_a \ga_b \ga_c = 
\fract{1}{3} \ga_a S_a \quad . 
\e \noindent
 For any $\g$ of dimension larger than 3, however, 
$K_3$ is not the only relevant fermionic $G$-invariant operator.
Indeed, for any $\g$ of rank $l$ we may introduce $l$ fermionic operators
by using its $l$ primitive cocycles. These cocycles are given
by skewsymmetric tensors of odd dimension $2m_i-1$ ($i=1,\dots,l$) 
where, for each $i$, $m_i$ is the order of the symmetric 
$G$-invariant polynomial giving the corresponding primitive Casimir-Racah 
operator. The structure constants used in (\ref{A4}) simply 
define the lowest order operator $K_3$, for which the ($m_1$=2)-order 
tensor is the Killing metric giving the quadratic Casimir. 
The higher order Casimirs and their associated ($2m_i-1$)-cocycles 
for a simple compact $\g$ are all known, 
and known to be relevant in many areas of physics,  
as in the mathematical description of anomalies
(see \cite{anom},~\cite{cup}),
current algebras and Schwinger terms 
(see {\it e.g.} \cite{fs}, \cite{dAim} and
references therein), Wess-Zumino terms
and effective actions (\cite{dHw},~\cite{dH},~\cite{dAmpb}),
$W$-algebras (\cite{bbs},~\cite{bosch}), principal chiral models 
(see \cite{ehmm} and earlier references therein), and others. 
For $\g=su(n)$, for example, 
the Casimir-Racah tensors of order $m_i=2,\dots,n-1$
give cocycles of orders $3,5,\dots ,(2n-1)$.
Writing $C_{abc}=f_{abc}$ for $su(n)$, 
the five-cocycle is determined by the third-order invariant 
tensor of coordinates $d_{abc}$ through
\be
\label{A6} \Omega_{abcde} =  f_{xa[b} f_{cd]y} d_{xye} \quad , 
\e \noindent
and is tabulated in ~\cite{tensors} for $n=3$ and $n=4$. 
Such a quantity can be seen 
{\it e.g.} as a Schwinger term in a two-dimensional chiral 
$SU(n) \times SU(n)$ model: see eq. (28) of \cite{dmg} 
and ~\cite{fs}.
Using (\ref{A6}) we may then construct, for some $k \in \R$
\be
\label{A5} 
K_5= -\fract{k}{5!}\Omega_{abcde} \ga_a \ga_b \ga_c \ga_d \ga_e\quad .
\e \noindent 
As we shall see, $K_5$ plays a non-trivial role in the 
understanding of the representation (\ref{A1}) already 
for $su(3)$, even though, in this case, $K_5$ is related to 
$K_3$ by Hodge duality as is proved in Sec. 3
and discussed in Sec. 6. 
For a given $\g$, higher order $K_{(2m_i-1)}$ operators
may be constructed similarly using the corresponding 
$\Omega_{(2m_i-1)}$ cocycles provided that the rank
$l$ is high enough. They all have odd character since they
are given by skewsymmetric tensors in the basis of the
$r$ different anticommuting $\gamma$'s.
 
To develop a deeper view, Sec. 4 is devoted to analyse 
the next smallest even case, that of $\g =su(5)$, for which 
there are four fermionic scalars $K_3, K_5, K_7, K_9$, one for 
each of the four primitive cocycles of $su(5)$. The higher 
cocycles give rise to operators which feature in any study of 
the system essentially on the same footing as the cubic 
Kostant operator (\ref{A4}). With the aid of some MAPLE 
programs, we are able to analyse various aspects of the role 
played by all the fermionic scalar operators $K_{(2m_i-1)}$.

  Our treatment will make this fermionic character explicit by
replacing $\gamma$ matrices, two at a time, 
as {\it e.g.} in (\ref{C2}) below, by 
Dirac fermions $A$ such that $\{ A \, , \, A^\dagger \} =1$. 
This is easy to do for
$r$ even. A similar approach to the $\mbox{dim}\,\g=r$ odd 
case is evidently complicated by the fact that it leaves over, 
unpaired, the last matrix $\gamma_r$ or, put otherwise, 
a single Majorana fermion type entity. This does not mean
that the odd case cannot be treated in a satisfactory 
way. It is however essential in treating the odd case to do so 
in a fashion that respects fully the fermionic nature of the 
unpartnered Majorana fermion, and also of the $K_3$ operator 
and the above generalisations. With this in mind, we 
treat in Sec.5 the ${\rm dim} \, \g=r=3$ case of $su(2)$ 
in detail to indicate how to handle the general 
odd case. In fact, when ${\rm dim} \, \g$ is odd, ${\cal D}$ 
involves the irrep of $\g$  with highest weight $\delta$ repeated 
in direct sum $2^{\fract{l-1}{2}}$ times.

We have referred to $K_3$ above as the Kostant operator. Strictly 
speaking, however, the operator $K$ recently introduced by Kostant  
~\cite{kos} (see also ~\cite{brra}) contains, besides the cubic
term $K_3$, an additional representation dependent piece that we 
ignore here by restricting our attention to the
purely geometrical part $K_3$ of $K$. Operators such as $K_3$,
however, are already familiar in non-relativistic supersymmetric 
quantum mechanics of particles with spin-$\fract{1}{2}$ 
(see {\it e.g.} ~\cite{dJ}) or colour 
degrees of freedom (see {\it e.g.} ~\cite{tani, ajmajm}). In such 
theories there are Majorana fermion variables $\psi_i$ with 
anticommutation relations 
\be 
\label{A7} \{ \psi_i \, , \, \psi_j \} = \delta_{ij} \quad , 
\e
\noindent 
for which there is a representation $\psi_i \mapsto \gamma_i / \sqrt{2}$ 
in terms of hermitian Dirac matrices. The supercharges of such
theories contain or consist of a term proportional to
\be 
\label{A9} 
iC_{ijk}\psi_i \psi_j \psi_k \quad . 
\e 
\noindent
In the case of $su(2)$, where $C_{ijk}=\epsilon_{ijk}$, then 
$S_i \mapsto \sigma_i /2$ describes spin one-half ~\cite{mar},~\cite{dJ}.
In this context, it may be remarked, because it underlies our 
interest, that more general models of particles with colour can possess
supercharges involving the higher fermionic scalars. Demonstrations of how
this arises, and its relationship to the mentioned hidden 
supersymmetries that do not close on the Hamiltonian of the model, 
will be addressed elsewhere.

\section{The representation ${\cal D}$}

It is easy to use (\ref{A2}) and (\ref{A3}) to show that
\bea
\label{B1} [S_a \, , \, \ga_b ] & = & iC_{abc} \ga_c \quad ,\nonumber \\
               {[}S_a \, , \, S_b {]} & = & iC_{abc} S_c \quad , 
\ea 

\noindent 
so that $X_a \mapsto S_a$ is indeed a representation of $\g$. 
Next we define the quadratic Casimir operator
\be
\label{B3} C_2=X_a \, X_a \quad . 
\e \noindent
Then using only (\ref{A3}) and the Jacobi identity for the 
structure constants of $\g$, it follows that for ${\cal D}$
\be
\label{B4} C_2({\cal D})= S_a \, S_a 
\e \noindent
has the form
\be 
C_2({\cal D})
= \fract{1}{8} c_2(ad).\mbox{dim}\,\g.{\bf 1}_{{\rm dim} \gamma} 
\quad , 
\label{B5}
\e \noindent
where $c_2(ad)$, the eigenvalue of $C_2$ for the 
adjoint representation $X_a \mapsto ad(X_a)$ given by
$(ad X_a)_{bc}=-iC_{abc}$, enters via
\be
\label{B6} {\rm Tr} \, (ad \, X_a adX_b) =c_2(ad) \delta_{ab} \quad . 
\e 
Thus, for $su(n)$, 
\be
C_2({\cal D})=\fract{1}{8} \, n(n^2-1).{\bf 1}_{{\rm dim} \,\gamma}\quad .
\label{casun}
\e 

   To identify the representation ${\cal D}$ in 
terms of the irreps of $\g$ is convenient to discuss
first the $r$ even case, even though the answer for
the odd-dimensional $\g$ case is very similar. 
Since the irreducible $\ga_a$ matrices of an $(r=2s)$-dimensional
space have dimension $2^s$, ${\cal D}$ must be at least of
dimension $2^s$ (larger if we used non-irreducible 
$\ga$'s). So we ask, what irreps of $\g$ 
are contained in  ${\cal D}$? The Weyl 
formula for the dimension of the irrep of
$\g$ with highest weight $\Lambda$ is given by the
Weyl formula (see, {\it e.g.}, ~\cite{slansky}, eq. (5.5)):
\be
\label{B7} N(\Lambda)= \prod_{positive{\phantom{x}}
 roots} (1+ {{(\Lambda,\alpha)}
\over {(\delta,\alpha)}}) \quad , 
\e 

\noindent
where the $\alpha$'s are the positive roots of $\g$. For the irrep 
corresponding to the Weyl principal vector $\Lambda=\delta
=(1,\mathop{\dots}\limits^l,1)$ we find, since there 
are $(r-l)/2$ positive roots, 
\be
\label{B8} N(\delta)= 2^{(r-l)/2} \quad . 
\e 

\noindent
Further this irrep does have the correct eigenvalue of the Casimir 
operator (\ref{B4}) to agree with (\ref{B6}). To see this we use 
the general result (see ~\cite{slansky}, eq. (5.10))

\be
\label{B9} c_2(\Lambda)= (\Lambda,\Lambda +2\delta) \quad , 
\quad C_2(\Lambda)=c_2(\Lambda).{\bf 1}_{{\rm dim}_\delta}\quad,
\e 

\noindent 
to find

\be
\label{B10}
c_2(\delta) =3 (\delta ,\delta) = \fract{1}{8} C_2(ad) \, r \quad 
\e 

\noindent
upon use of the `strange formula' of Freudenthal and de Vries
(see ~\cite{FuchsSchw}, eqs. (7.20) and (14.30)). For $su(n)$, 
(\ref{B10}) reproduces the scalar in (\ref{casun}) 
since $c_2(ad)=n$. 
Moreover, we see from eq.(\ref{B5}) that, for any $\g$, 
$c_2(\delta)=c_2({\cal D})$. Since, for $r$ even, the irreducible 
gammas that determine the dimension of ${\cal D}$ are $p \times p$ 
matrices with $p=2^{{r \over 2}}$,  and the irrep 
$(1,\mathop{\dots}\limits^l,1)$ with the 
correct value for its Casimir operator involves $q \times q$ 
matrices with $q=2^{\fract{r-l}{2}}$, the natural
conclusion is that ${\cal D}$ is the direct sum of exactly 
$2^{\fract{l}{2}}$ copies of $(1,\mathop{\dots}\limits^l,1)$. 
When $\g$ is odd-dimensional, $r=2s+1$, ${\cal D}$ is 
of dimension $2^s=2^{\fract{r-1}{2}}$ and the same reasoning shows that
the irrep of highest weight $\Lambda$=$\delta$, of dimension
$2^{\fract{r-l}{2}}$ is contained $2^{\fract{l-1}{2}}$ times
in ${\cal D}$.

We may see this explicitly by
constructing the required number of copies of the irrep 
$(1,\mathop{\dots}\limits^l,1)$ in a fermionic 
Fock space description of ${\cal D}$. 
We first consider the even case $\g=su(3)$ fully; this
not only shows how our realisation works in the simplest non-trivial 
case but also indicates exactly how the general even 
case is handled. The odd $r$ case will be discussed in Sec. 5.

\section{The fermionic Fock space ${\cal V}_{\cal D}$ for $su(3)$}

We build the fermionic Fock space of ${\cal D}$ for $su(3)$ by
constructing four Dirac fermions out of the eight available gamma
matrices $\ga_a$, 
where the index $a=1,\dots,8$ also 
labels the $su(3)$ generators (say, the Gell-Mann $\lambda_a$ 
matrices, $[\lambda_a,\lambda_b]=2if_{abc}\lambda_c$). However 
not all four Dirac fermions enter on the same footing, nor 
do they have the same role to play. So, on the one hand, we set

\be
\label{C1} 2B=\ga_3+i\ga_8 \quad , \quad 2B^\dagger=\ga_3-i\ga_8 \quad , 
\e

\noindent 
corresponding to the $su(3)$ Cartan subalgebra 
generators $\lambda_3$ and $\lambda_8$, and, on the other, 
corresponding to the positive roots, we define

\be
\label{C2} 2A_5=\ga_4-i\ga_5 \quad , \quad  2A_2=\ga_6+i\ga_7  
\quad , \quad  2A_1=\ga_1+i\ga_2 
\e

\noindent  
(the somewhat strange labelling is adapted to the $su(5)$ case in
next section). In this way, (\ref{A3}) translates into
\bea
\label{C3} \{ A_i \, , \, {A_j}^\dagger \} & = & \delta_{ij} \quad , \quad
\{ A_i \, , \, A_j \}=0 \quad , \nonumber \\
\{B \, , \, B^\dagger \} & = & 1 \quad , \quad B^2=0 \quad , \quad
\{B, A_j \} =0 \quad , 
\ea 

\noindent 
which exhibit the fermionic nature of our realisation. Defining 
the fermion number operators

\be
N_i={A_i}^\dagger\, A_i\quad (i=5,2,1) \quad,\quad N_B=B^\dagger \,B\quad , 
\label{C4}
\e
\be
\gamma_4\gamma_5=i(2N_5-1)\;,\;\gamma_6\gamma_7=-i(2N_2-1)\;,
\; \gamma_1\gamma_2=-i(2N_1-1)\; ,\; \gamma_3\gamma_8=-i(2N_B-1)
\label{N}
\e

\noindent 
we find 

\be
\label{C5} I_z=S_3=\fract{1}{2} (N_2+N_5-2N_1) \quad , 
\e

\be
\label{C6} Y = \fract{{2}}{{\sqrt{3}}} S_8 = N_5 -N_2 \quad, 
\e

\noindent 
independently of $N_B$. 

It is now easy to observe the correspondence
between the weights of an $su(3)$ octet (irrep $(1,1)=\{8\}$) and 
fermionic states labelled by occupation numbers
\bea
\label{C7}
\qquad |n_5,n_2,n_1 ) & \quad & (I_z \, , \, Y) \nonumber \\
|0,0,0 ) & \quad  & (0,0) \nonumber \\
|1,0,0 ) & \quad & (\fract{1}{2},1) \nonumber \\
|0,1,0 ) & \quad & (\fract{1}{2},-1) \nonumber \\
|0,0,1 ) & \quad & (-1,0) \nonumber \\
|0,1,1 ) & \quad & (-\fract{1}{2},-1) \nonumber \\
|1,0,1 ) & \quad & (-\fract{1}{2},1) \nonumber \\
|1,1,0 ) & \quad & (1,0) \nonumber \\
|1,1,1 ) & \quad & (0,0) \quad . \ea 

\noindent 
Thus, apart from the eigenvalue $I$ of the
$su(2)$ Casimir, which is used to distinguish the octet central 
states, the 
$|n_5,n_2,n_1) $ part of the full Fermi Fock space determines
the bulk of the expected $su(3)$ characteristics. To complete the
characterization we may use the 
$su(3)$ standard $I,\; U,\; V$-spin raising and lowering operators. 
These are 

\bea
\label{C8} I_-=S_1-iS_2 & = & -(B+B^\dagger) {A_1}^\dagger +A_2 \, A_5 
\quad , \nonumber \\
V_-=S_4-iS_5 & = & (\omega B +\omega^2 B^\dagger) A_5+{A_2}^\dagger \,
{A_1}^\dagger \quad , \nonumber \\
U_+=S_6+iS_7 & = & (\omega^2 B +\omega B^\dagger) A_2 + 
{A_1}^\dagger{A_5}^\dagger \quad , 
\ea 

\noindent
where 
$\omega =-\fract{1}{2} (1-i\sqrt{3})$, $\omega^3=1$, and 
$I_+=I^\dagger_-\,,\, V_+=V_-^\dagger\,,\, U_-=U_+^\dagger\,$. 
The way $B$ enters here points the way directly 
towards the construction of two sets of orthogonal octet states
in the fermionic Fock space of ${\cal D}$. We write 
$|n_B, n_5, n_2, n_1 \rangle$ to denote the 
${\cal V}_{\cal D}$ Fock space states that are simultaneous eigenstates of the 
number operators $N_B$ and the $N_i$; the standard $su(3)$ octet 
states of chirality $\pm$ are labelled $|8, \pm, I, I_z, Y\rangle$. 
We define chirality by 

\be
\label{C9} \ga_9 \equiv -\prod_{a=1}^8 \ga_a = 
(2N_B-1) \prod_{k=5,2,1} (2N_k-1)
=(-1)^{N_B+N_5+N_2+N_1}=(-1)^{tot. Fermi \#} \quad , 
\e 

\noindent     
which commutes with the generators $S_a$ of the representation 
${\cal D}$ of $su(3)$. To complete the construction  we label the 
states of the two octets of different chiralities 
$|8,\pm,I\, I_z \, Y \rangle$. To relate these to the Fock
 $|n_B, n_5, n_2, n_1 \rangle$ states we refer 
to (\ref{C7}), and identify the highest weight 
$\pm$ octet states with the Fock space states of the correct 
chirality using

\be
\label{C10}
|8,\pm,I=1,I_z=1 \, Y=0 \rangle \equiv 
|n_B= {0 \atop 1},n_5=1,n_2=1,n_1=0\rangle \quad . 
\e 

\noindent
The other octet states are constructed using $V_-$ and $U_+$ 
each once and then $I_-$ as needed to get all other states 
except the central state with $I=I_z=Y=0$. 
One gets this last state by orthogonality with the state with
$I=1 \,, I_z=Y=0$. The results 
that give the states $|8,\pm,I,I_z,Y \rangle$ in terms of
$|n_B,n_5,n_2,n_1\rangle$ with the correct phases thus are
\bea
\label{C11}
|8,\pm,\fract{1}{2}, \fract{1}{2}, 1\rangle &=
& U_+| \tatop{0}{1} ,1,1,0\rangle = 
\tatop{{-\omega}}{{\omega^2}} |\tatop{1}{0} ,1,0,0 \rangle \; , 
\nonumber \\
|8,\pm,\fract{1}{2}, \fract{1}{2}, -1\rangle &=
& V_-|\tatop{0}{1} ,1,1,0 \rangle = 
\tatop{{\omega^2}}{{-\omega}} |\tatop{1}{0} ,0,1,0 \rangle \; , 
\nonumber \\
|8,\pm,\fract{1}{2}, -\fract{1}{2}, 1\rangle &=
& I_-|8,\pm,\fract{1}{2},\fract{1}{2} \, 1 \rangle= 
\tatop{{\omega}}{{\omega^2}} | \tatop{0}{1} ,1,0,1 \rangle \; , 
\nonumber \\
|8,\pm,\fract{1}{2}, -\fract{1}{2}, -1\rangle &=
& I_-|8,\pm,\fract{1}{2} \,, \fract{1}{2} \,, -1 \rangle= 
\tatop{{-\omega^2}}{{-\omega}} | \tatop{0}{1} ,0,1,1 \rangle \; , 
\nonumber \\
|8,\pm,1 \,,0\,0\rangle  &=
& \fract{1}{{\sqrt{2}}} I_-|\tatop{0}{1} ,1,1,0\rangle=
\fract{1}{{\sqrt{2}}} \left(\mp \, | \tatop{1}{0} ,1,1,1\rangle 
+|\tatop{0}{1} ,0,0,0 \rangle \right) \; , 
\nonumber \\
|8,\pm,1,-1,0 \rangle &=& \fract{1}{{\sqrt{2}}} I_-
|8,\pm,1,0,0 \rangle= \pm |\tatop{1}{0} ,0,0,1 \rangle \quad 
\ea

\noindent 
together with

\be
\label{zerost}
|8,\pm,0,0,0 \rangle=
\fract{1}{{\sqrt{2}}}|\tatop{1}{0},1,1,1\rangle
\pm |\tatop{0}{1} ,0,0,0 \rangle \quad .
\e
Since the $I,\; U$ and $V$ spin operators are all even, 
as any generator should be, all states within a given octet have 
the same chirality by (\ref{C9}). Thus, chirality 
distinguishes the two octets 
$\{8,\pm\}$ of even/odd total fermion number that span ${\cal D}$ 
in this case.

Let us now look for the role played by the two fermionic
scalars $K_3$, $K_5$. For $K_3$ we may use the results given so 
far to find its complete operator expression in the full fermionic 
Fock space ${\cal V}_{\cal D}$. This gives 
\be
\label{C12} K_3=L_{125}-B(N_1+\omega N_5+\omega^2 N_2)
    -B^\dagger(N_1+\omega^2 N_5+\omega N_2) \quad , 
\e
\noindent 
where $L_{125}= A_1 \, A_2 \, A_5+ {A_5}^\dagger \,
{A_2}^\dagger \, {A_1}^\dagger$. $L_{125}$ in eq. (\ref{C12})
affects only the two central octet states, while
the other terms affect only the states of the hexagonal 
rim of the octet ({\it i.e.} the states of the orbit of the state (\ref{C10})
under the Weyl group of $su(3)$).
Eq. (\ref{C11}) leads easily to the result
\be
\label{C13} K_3 |8,\pm,I \, I_z \, Y \rangle 
=|8,\mp,I \, I_z \, Y \rangle \quad . 
\e 

\noindent 
Thus, $K_3$ changes the chirality of the state while respecting
the $su(3)$ labels. This does not depend on $\g$ being $su(3)$: 
it is a consequence of $K_3$ being realized by a 
three-cocycle of a $\g$, since a cocycle is both odd (so that 
$\{K_3,\gamma_9\}=0$) and $G$-invariant. Thus, similar remarks apply also
for the higher order fermionic operators $K_{(2m_i-1)}$ 
associated with the various primitive cocycles of any $\g$, since they 
are all odd and $G$-invariant and, in particular, to $K_5$ for $su(3)$.

Since for larger $\g$ and higher cocycles the task of finding complete 
operator expressions for quantities like $K_3,\; K_5,\dots$
becomes rapidly more time-consuming, it is better to seek 
expressions only for their parts that are non-trivial in their action 
on highest weight states. We illustrate this for $K_3$ first. 
We have 
\be \label{C14}
 K_3=-{1 \over {12}} if_{abc} \gamma_a \gamma_b \gamma_c=-{1 \over 2} i
\sum_{triples} f_{abc} \gamma_a \gamma_b \gamma_c \quad , 
\e \noindent over non-trivial triples such that $a < b < c$, so that
\bea
K_3 & = & -{1 \over 2} i\{ \gamma_{123}+{1 \over 2}( \gamma_{147}+ 
\gamma_{165}+  \gamma_{257}+ \gamma_{246})+{1 \over 2}(\gamma_{345}
+\gamma_{376})+ {{\sqrt{3}} \over 2}(\gamma_{458}+\gamma_{678})\}  
\nonumber \\
 & = & {1 \over 2}(N_2+N_5-2N_1)\gamma_3 +{{\sqrt{3}} \over 2}
(N_5-N_2)\gamma_8 +\dots
\nonumber \\
 & = &  \gamma_3 S_3+
\gamma_8 S_8 + \dots \quad  , \label{C15} \ea
\noindent 
by (\ref{C5}), (\ref{C6}). Here the dots indicate terms 
(coming from the second bracket of (\ref{C15})) that give vanishing 
contribution to the action of $K_3$ on the Weyl orbit of (\ref{C10}).
Hence on (\ref{C10})
\be \label{C17}
K_3 = \gamma_3 =B+B^{\dagger}\quad {\rm and} \quad K_3^2=1 \quad , 
\e \noindent in agreement with (\ref{C12}).

Similarly, for actions on (\ref{C10}) we have
\be \label{C18}
K_5=-{k \over{5!}} \Omega_{abcde} \gamma_a \gamma_b \gamma_c 
\gamma_d \gamma_e =-k\sum_{pentuples} 
\Omega_{abcde} \gamma_a \gamma_b \gamma_c \gamma_d \gamma_e \quad ,
\e \noindent 
over pentuples such that $a \in \{ 3,8 \}, \; c=b+1, \; c<d, \;
e=d+1$. This is justified: the total antisymmetry of $\Omega_{sabcd}$ allows 
us to order the indices in any convenient way and sum over pentuples that 
both respect this order and give terms that fail to annihilate (\ref{C10}).
We find
\be \label{C19}
K_5=k\gamma_3(\Omega_{31245}-\Omega_{31267})
 +k\gamma_8(\Omega_{81245}-\Omega_{81267}-\Omega_{84567})
=k {{\sqrt{3}} \over {12}} \gamma_8(1+1+2) \quad , 
\e \noindent 
using data from table 3 of ~\cite{tensors} and the results 
\be 
\label{C20}
i\gamma_1 \gamma_2=2N_1-1 \to -1 \; , \;
i\gamma_4 \gamma_5=1-2N_5 \to -1 \; , \;
i\gamma_6 \gamma_7=2N_2-1 \to 1 \quad ,
\e \noindent 
for actions on (\ref{C10}). Thus
\be 
\label{C21}
K_5=\frac{k}{\sqrt{3}}\,\gamma_8 \quad {\rm and} \quad K_5^2=
\fract{1}{3} \, k^2 
\e 
\noindent 
on (\ref{C10}). It was also checked explicitly that non-zero 
coefficients of the type $\Omega_{38cde}$ do not give rise to 
any non-vanishing contributions to (\ref{C21}). 

Thus, setting $k=\sqrt{3}$, and acting on the vector space 
spanned by the two highest weight vectors of (\ref{C10}), 
\be 
\label{C22}
|n_b=0, HW \rangle \quad , \quad |n_b=1 , HW \rangle  \quad , 
\e 
\noindent 
we find
$K_3 \mapsto \sigma_1$, and  $K_5 \mapsto \sigma_2$, where
$2B=\gamma_3 +i \gamma_8$. 
Hence $\chi=-iK_3 K_5 \mapsto -i\sigma_1 \sigma_2
=\sigma_3$. We may also use Hodge duality to give a direct proof
\bea 
K_5 & = & 
(-{{\sqrt{3}} \over {5!}}) \Omega_{pqrst} \gamma_{pqrst} \nonumber \\
& = &  \left({{-1} \over {12.5!}} \right) \gamma_{[pqrst]}
\epsilon_{pqrstxyz} f_{xyz}  \nonumber \\
& = & {1 \over {12}} f_{xyz} \gamma_{[xyz]}\gamma_9 \;=\;i K_3 \gamma_9 
\quad .
\label{C23} 
\ea 
\noindent Here, we have used Eq.(8.14) of ~\cite{tensors}, the definition
(\ref{C9}) of $\gamma_9$ and the result
\be \label{C24} 
\epsilon_{pqrstxyz}  \gamma_{[pqrst]}=- 5! \, \gamma_{[xyz]} \gamma_9 \quad .
\e 
\noindent Eq, (\ref{C23}) implies
\be \label{C23A}
-iK_3 K_5=\gamma_9 \e \noindent
known to be represented by $\sigma_3$.
  As a final remark on the $su(3)$ case, we 
notice that insertion of (\ref{C5}) to (\ref{C8}) into (\ref{B4}) 
gives rise to the result
\be
\label{C25} C_2({\cal D})=3\,{\bf 1}_{16} \quad , 
\e \noindent as it should by eq. ({\ref{casun}), upon cancellation of all 
number operator terms.

\section{The general $\g$ even case}
\subsection{General remarks} 

The method of Sec. 3 extends directly to a general 
even-dimensional $\g$ and indeed, without much modification, to the 
odd case (Sec. 5). For $r$ ($l$) even, we define
$\fract{l}{2}$ operators $B_1, \dots ,B_{l/2}$, and 
$(r-l)/2$ operators $A_1, \dots ,A_{(r-l)/2}$, their adjoints and the 
corresponding number operators $N_{B_\mu}$ and $N_{A_i}$. 
To label the states within an irrep one needs the $l$ labels 
provided by the Cartan subalgebra generators plus
$(r-3l)/2$ additional ones to sort out the possible degeneracy.
Since ${\cal D}$ contains $(1,\mathop{\dots}\limits^l,1)$ 
$2^{l/2}$ times (Sec. 2), we still need $l/2$ labels taking 
two possible values to distinguish the states of the 
different copies of $(1,\mathop{\dots}\limits^l,1)$ 
in ${\cal D}$. This means, in all, $r/2$ 
labels, provided by the $l/2$ operators $N_{B_\mu}$ and the $(r-l)/2$
operators $N_{A_i}$. There is thus one $A$ for each of the 
positive roots and we can establish of a one to one 
correspondence like (\ref{C7}) between the weights of the irrep 
$(1,\mathop{\dots}\limits^l,1)$ of $\g$ and the simultaneous 
eigenstates of the commuting number operators $N_{A_i}$.
As seen in already for the $su(3)$ case (eq.(\ref{C8})), 
however, {\it both} the $B_\mu$'s and the $A_i$'s appear
in the definition of the various ladder operators that
generate the states of the $(1,\mathop{\dots}\limits^l,1)$
representation. The eigenvalues 
$n_1,\dots ,n_{l/2}$ of the $N_{B_\mu}$ 
can be used as in (\ref{B10}) to define the Fock space states 
equal to the highest weight states 
($2^{\fract{l}{2}}$ of them)
of the different copies of $(1,\mathop{\dots}\limits^l,1)$ 
in ${\cal D}$. If one builds the rest of the states of each 
copy by application of lowering operators, one will find that 
any one state of any copy is orthogonal to all states of any 
other copy, as well, of course, to all of the states of 
its own copy. 

The details of the next simplest even $\g$ of higher rank, 
$su(5)$, offers further insight into the situation 
surrounding the fermionic $SU(5)$-invariant operators 
$K_3, \; K_5, \; K_7, \; K_9$ built with the aid of its four non-trivial 
cocycles. Our discussion for $su(5)$ proceeds along lines similar 
to those followed for $su(3)$.
 
\subsection{Basic definitions for $su(5)$}

We use an explicit and essentially standard ({\it cf.} 
~\cite{hayashi}) set of Gell-Mann lambda matrices for $su(5)$. 
For the diagonal matrices $\lambda_{(p^2-1)},\, (p=2,3,4,5),$ 
we have
\bea
\lambda_3={\rm diag} \, (1,-1,0,0,0) \quad & , & \quad \sqrt{3} \lambda_8=
{\rm diag} \, (1,1,-2,0,0) \quad , \nonumber \\
\sqrt{6}\lambda_{15}={\rm diag} \, (1,1,1,-3,0,) \quad & , & \quad 
\sqrt{10}\lambda_{24}={\rm diag} \, (1,1,1,1,-4) \quad . \label{D1} \ea
\noindent  The index pairs 
\be \label{D1A}
(1,2), (4,5), (6,7), (9,10), (11,12), (13,14), (16,17), (18,19), (20,21),
(22,23) \quad, \e \noindent
are associated with the remaining $\lambda$'s in a way that can be 
inferred from the following array:
\be \label{D2}
\left( \begin{array}{ccccc}
- & (1,2) & (4,5) & (9,10) & (16,17) \\
- & - & (6,7) & (11,12) & (18,19) \\
- & - &  -& (13,14) &(20,21) \\
- & - & - & - & (22,23) \\
- & - & - & - & - \end{array} \right)  \quad .
\e \noindent 
For example, $\lambda_6$ is a symmetric matrix whose only 
non-zero element above the main diagonal is a $1$ at the place 
marked $(6,7)$ in (\ref{D2}), while $\lambda_7$ is antisymmetric 
with a single entry $-i$ in the same place.
Next, follow the succesive `diagonals' of the array to
define the $su(5)$ Dirac fermions for the various pairs as follows.
\bea 
2A_1=\gamma_1 +i\gamma_2 \quad & , &  \quad 2A_2=\gamma_6+i\gamma_7 \quad ,
\nonumber \\
2A_3=\gamma_{13} -i\gamma_{14} \quad & , & \quad 
2A_4=\gamma_{22} +i\gamma_{23} \quad , \nonumber \\
2A_5=\gamma_4 -i\gamma_5 \quad & , & \quad 
2A_6=\gamma_{11} +i\gamma_{12} \quad  , \nonumber \\
2A_7=\gamma_{20} +i\gamma_{21} \quad & , &  \quad 
2A_8=\gamma_9 +i\gamma_{10} \quad , \nonumber \\
2A_9=\gamma_{18} -i\gamma_{19} \quad & , &  \quad 
2A_{10}=\gamma_{16} -i\gamma_{17} \quad , 
\label{D3}
\ea 
\noindent 
plus 
\be
\label{su5bes}
2B_1=\gamma_3 +i\gamma_8 \quad , \quad  2B_2=\gamma_{15} -i\gamma_{24}
\quad.
\e
This leads to relations with the number operators $N_{\alpha}=
{A_{\alpha}}^\dagger A_{\alpha}$, like
\bea
i \gamma_1 \gamma_2 =2N_1-1 \quad {\rm and\; similarly\; for} \quad 
& N_\alpha & {\rm when} \quad \alpha \in \{1,2,4,6,7,8 \} \nonumber \\
-i \gamma_4 \gamma_5 =2N_5-1 \quad {\rm and\; similarly\; for} \quad  
& N_\alpha & {\rm when} \quad  \alpha \in \{3,5,9,10 \} \quad .
\label{D4} 
\ea \noindent 
Then $S_a=-{1 \over 4}i f_{abc} \gamma_b \gamma_c$, with the aid of 
MAPLE output for structure constants, allows us to derive
\bea
2S_3\quad = \quad 2I_3 & = & N_2+N_5-2N_1-N_8+N_{10}-N_9+N_6 \quad , 
\nonumber \\
S_8= {{\sqrt{3}} \over 2} Y \quad , \quad 3Y & = 
& (3N_5-3N_2-N_8-N_6+N_9+N_{10} -2N_3+2N_7) \quad , 
\nonumber \\
S_{15}={{\sqrt{6}} \over 3} Z_3 \quad , \quad 4Z_3 & = 
& (4N_3-4N_6-4N_8+N_9+N_{10}-N_7 +3N_4) \quad , 
\nonumber \\
S_{24}={{\sqrt{10}} \over 4} Z_4 \quad , \quad Z_4 & = 
& N_9+N_{10}-N_4-N_7 \quad ,
\label{D5} \ea
\noindent 
which reproduce the $su(3)$ expressions when only the 
(1,2 and 5)-labelled quantities are retained. Our choice of signs 
in (\ref{D3}) depends upon the signs of the structure constants for 
$su(5)$ -- our choice of lambda-matrices was made above to yield agreement 
with the tables given in \cite{hayashi}, 
upon our wish of avoiding constant terms in the definitions (\ref{D5}) 
and upon our desire of having all entries in (\ref{D7A}) below 
equal to +1.

We  define the highest weight state of any irrep of $su(5)$ by 
taking first the highest $Z_4$ eigenvalue, then the highest $Z_3$ 
eigenvalue that can arise for that $Z_4$ eigenvalue. Next the highest 
$Y$ and finally the highest $I_3$. Thus we get $Z_4=2$ for 
$N_4=N_7=0 , N_9=N_{10}=1$. Hence $Z_3=-N_8-N_6+N_3+{1 \over 2} 
={3 \over 2}$ for $N_8=N_6=0,N_3=1$, and $Y=N_5-N_2=1$ for $N_2=0,N_5=1$. 
Finally $I_3={1 \over 2}$ for $N_1=0$. Hence our highest weight 
state for any of the four possible irreps $(1,1,1,1)$ of $su(5)$ 
in ${\cal D}$ 
includes the $(r-l)/2=10$ labels 
\be 
\label{D6} 
|0,0,1,0,1,0,0,0,1,1 ) \quad , 
\e
\noindent 
where we have written the eigenvalues $n_{\alpha}$ of the 
$N_{\alpha}$ with the  $\alpha$ in the standard 
order $1, \dots ,10$. We remark that our method of definition corresponds to 
the use of the subgroup chain in which $(1,1,1,1)$ of $su(5)$ reduces to 
$(1,1,1)_2$ of $su(4) \times u(1)$, plus irreps of lower $Z_4$, and then
$(1,1,1)$ of $su(4)$ reduces to $(1,1)_{{3 \over 2}}$ of $su(3) \times u(1)$,
plus irreps of $su(3)$ of lower $Z_3$, and so on.
  
For $\g=su(5)$, there are two operators $N_{B_1}$ and $N_{B_2}$,
providing the $l/2=2$ additional labels $|n_{B_1}, n_{B_2})$
that give rise to the possibilities
\be
\label{D6A} |0,0) \, ,\, |1,1) \, , \, |0,1) \, ,\, |1,0) \quad , 
\e \noindent
available to construct the highest weight states of chiralities $+,+,-,-$
of the four different copies of the $(1,1,1,1)$ irrep contained in 
${\cal D}$. The complete irreps can be built from these by lowerings. 

The $K_3$ operator commutes with all $su(5)$ actions 
and changes chiralities. This does not furnish a complete 
picture; there are three other fermionic scalar operators 
that deserve to be treated on the same footing as the simplest 
one $K_3$. We turn to them next.

\subsection{$su(5)$ fermionic operators}

We define the 5th order fermionic operator for $su(5)$ 
via the five-cocycle $\Omega_{sabcd}$ by
\bea
K_5 & = & -{{k_5} \over {5!}} \Omega_{sabcd} \gamma_s \gamma_a 
\gamma_b \gamma_c \gamma_d \quad, \quad k_5 \in R \quad , \nonumber \\
& = & k_5 \sum_{pentuples} \Omega_{sabcd} (-i \gamma_a \gamma_b)
(-i \gamma_c \gamma_d) \gamma_s \quad , \label{D7} \ea \noindent
over pentuples such that $s \in \{ 3,8,15,24 \}$, and  $a \in 
 \{ 1,4,6,9,11,13,16,18,20 \}$, and
$b=a+1 \; , \; c>b \; , \;
d=b+1$. These are the only ones that can fail to annihilate the 
highest weight state. In fact, for the action of $K_5$ on (\ref{D6}), 
it follows that 
\be \label{Dinsert}
(-i\gamma_x \gamma_{x+1})=1 \quad \mbox{for all}\quad 
x \in \{ 1,4,6,9,11,13,16,18,20,22 \} \quad . \e \noindent
Thus if define an array 
$N_{ab}$ whose only non-vanishing entries 
on (\ref{D6}) are 
\be 
\label{D7A} 
N_{12}=N_{45}=N_{67}=
N_{9,10}=N_{11,12}=N_{13,14}=N_{16,17}=N_{18,19}=N_{20,21}=N_{22,23}=+1, 
\e
we may do a computation of the coefficients in $K_5$ 
of the $\gamma_s$, for each of  $s \in \{3,8,15,24 \}$, by a separate 
MAPLE run. This yields a result valid for the action of $K_5$ on highest 
weight states:
\be 
\label{D8} 
2K_5=k_5(3\gamma_3+{5 \over 3} \sqrt{3} \gamma_8 +{1 \over 3} \sqrt{6} 
\gamma_{15} -\sqrt{10} \gamma_{24}) \quad . \e
\noindent The same procedure works for the 
7th order fermionic scalar $K_7$ of $su(5)$, given by
\bea 
K_7 & = & i{{k_7} \over {7!}}\Omega_{sabcdef}
 (-i\gamma_{ab}) (-i\gamma_{cd}) (-i\gamma_{ef}) \, \gamma_s \nonumber \\
 & = & k_7 \sum_{heptuples} \Omega_{sabcdefg}
 (-i\gamma_{ab}) (-i\gamma_{cd}) (-i\gamma_{ef}) \,\gamma_s \quad ,\label{D9}
\ea \noindent over heptuples with $b=a+1,d=c+1,f=e+1,c>b,e>d$ and 
$$a \in \{ 1,4,6,9,11,13,16,18 \} \quad .$$ 
 
\noindent This leads, by four runs of the corresponding
MAPLE program, to 
\be \label{D10}
K_7=k_7({2 \over 5} \gamma_3-{2 \over {45}} \sqrt{3} \gamma_8 -
{7 \over {45}} \sqrt{6} \gamma_{15} +{1 \over {15}} \sqrt{10} \gamma_{24}) 
\e 
on the HW states (\ref{D6}).

It is of course possible to derive (by MAPLE program) that
\be \label{D11}
2K_3=\gamma_3+\sqrt{3} \gamma_8 +\sqrt{6} \gamma_{15} +\sqrt{10} \gamma_{24}
\e
\noindent 
on (\ref{D6}). But it is easier to derive this directly from
\be \label{D12}
K_3=S_3 \gamma_3+S_8 \gamma_8+S_{15} \gamma_{15}+S_{24} \gamma_{24} 
\quad , \e
\noindent and insertion of the known eigenvalues of the $S_s$ 
for (\ref{D6}) gives back (\ref{D11}).
To see that (\ref{D12}) for $K_3$ is correct, 
think in terms of triples $s,x,x+1$ with $s \in \{ 3,8,15,24 \}$ and
\be \label{D13}
x \in \{ 1,4,6,9,11,13,16,18,20,22 \} \quad . \e
\noindent The
relevant $f_{abc}$ are in the tables of ~\cite{hayashi} and we used data from 
MAPLE, which agreed with these tables.

\subsection{{\bf Discussion of $su(5)$ results}}

By giving a direct evaluation that uses Jacobi identites, it may be shown
that the different fermionic scalars $K_3$, etc.,
anticommute. It is wise to verify that the results (\ref{D8}), (\ref{D10}), 
(\ref{D13}) are in agreement with this. This requires only that the 
numbers that come from terms like ${\gamma_3}^2=1$, etc. sum to zero, 
which they indeed do. Our results also give the squares of the fermionic 
scalars
\be 
\label{D14}
{K_3}^2=5 \; , \; {K_5}^2=7{k_5}^2 \; , \; {K_7}^2={{16} \over {45}} 
{k_7}^2  \quad . 
\e 
\noindent
Fortunately we know sufficient cocycle identities 
valid for all $su(n)$ to complete a worthwhile check 
upon our MAPLE computation methods.

We have $K_3=-{1 \over {12}} if_{abc} \gamma_a \gamma_b \gamma_c$, and also 
${K_3}^2 =xI$ for some $x \in \R$, since ${K_3}^2$ is a scalar. Hence
\be 
\label{D15}
({\rm dim} \, \gamma ) x= {\rm tr}{K_3}^2= {1 \over {24}} ({\rm dim} \, 
\gamma ) f_{abc} f_{abc} \quad , 
\e
\noindent 
using Dirac trace methods with ${\rm tr} \, I={\rm dim} \, \gamma$. 
Since $f_{abc} f_{abc}=n(n^2-1)=5.24$, we find $x=5$, as 
required. 
To confirm the answers (\ref{D14}) for
the cases of $K_5$ and $K_7$, with $k_5$ and $k_7$ set equal to $1$, 
we need the identities which are the $n=5$
special cases of
\be \label{D16}
\Omega_{abcde} \Omega_{abcde}={1 \over 3} n(n^2-1)(n^2-4) \quad , \e
\be \label{D17}
\Omega_{abcdefg} \Omega_{abcdefg}={2 \over {45}}n(n^2-1)(n^2-4)(n^2-9)
\quad . \e
\noindent Thus, setting ${K_5}^2 =yI$ for some $y \in \R$, we obtain
\be \label{Dextra}
({\rm dim} \, \gamma ) y= {\rm tr}{K_5}^2={1 \over {5!}}
\Omega_{abcde} \Omega_{abcde} ({\rm dim} \, \gamma ) \e 
\noindent upon evaluating the trace in a fashion that takes full advantage of 
antisymmetries. Hence, using (\ref{D16}) at $n=5$, we find $y=7$, as expected.
 
We note the right hand sides of (\ref{D16}) and (\ref{D17})    
are zero for low enough $n$ as consistency requires.
We note also that (\ref{D16}) evaluated for $n=3$ allows us to see 
agreement with the result (\ref{C21}), when one puts $k=1$ in (\ref{C21}).
Also, we check from (\ref{D11}),(\ref{D8}), and (\ref{D10}) that
$K_3,K_5$ and $K_7$ anticommute with each other. Next, by asking 
for the unique linear combination that anticommutes with
$K_3,K_5$ and $K_7$, we can show that
on highest weight states (\ref{D6}) 
\be 
\label{D18}
K_9=k_9 \, (10\gamma_3-10\sqrt{3}\gamma_8+5\sqrt{6} \gamma_{15}-
\sqrt{10} \gamma_{24}) \quad , 
\e 
\noindent 
so that 
\be 
\label{D19} 
{K_9}^2=560 {k_9}^2 \quad . 
\e

Setting $k_5, k_7, k_9 =1$, we can compute the effect on (\ref{D6}) 
of the product $ K_3 K_5 K_7 K_9$. It is found that, as expected, 
only terms containing permutations of 
$\gamma_3 \gamma_8 \gamma_{15} \gamma_{24}$ survive, so that
\be 
\label{D20}
K_3 K_5 K_7 K_9={{16} \over 3}.7.\sqrt{5}.\gamma_3 \gamma_8 
\gamma_{15} \gamma_{24} \quad . 
\e
\noindent 
If we define $L_3$ so that $L_3=c_3 K_3$ and $L_3^2=1$, and so on,
then above results allow the choice
\be \label{D21}
c_3=\sqrt{5} \; , \; c_5=\sqrt{7} \; , \; c_7= {3 \over 4} \, 
\sqrt{5} \; , \; c_9= {1 \over {4 \sqrt{35}}}  \quad , \e
\noindent so that 
\be 
\label{D22}
\chi = L_3 L_5 L_7 L_9=\gamma_3 \gamma_8 \gamma_{15} \gamma_{24} 
\quad , \e
\noindent and $\chi^2=1$. 

   In view of the direct computation (\ref{C23}) for $su(3)$ and
of (\ref{D20}), we might have expected to find
\be 
\label{Dgamma25}
\chi = \gamma_{25} = - \prod_{\alpha=1}^{24} \gamma_{\alpha} 
\e
\noindent 
rather than (\ref{D22}) for (factors apart) the product 
of the $K$'s in (\ref{D20}). However, recalling from (\ref{Dinsert}) 
that for action upon (\ref{D6}) $\quad \gamma_1 \gamma_2=i$ etc.,
we see that all the $\gamma$'s absent from (\ref{D22}) can be 
smuggled back (\ref{D22}) at the sole cost of a factor $i^{10}=-1$, 
so that the two expressions for $\chi$ (\ref{D22}), (\ref{Dgamma25})
coincide. We shall come back to this point in Sec. 6.

Since the $L$ operators commute with the $su(5)$ action, we can 
consider the effect on the vector of $su(5)$ highest weight states
\be 
\label{D23}
|00,HW\rangle \; , \; |11,HW\rangle \; , \; |01,HW\rangle \; , \;
|10,HW\rangle \quad , 
\e
\noindent 
where the first two labels provide the eigenvalues of the number
operators $N_{B1}$ and $N_{B2}$, associated with the Dirac fermions
$2B_1$ and $2B_2$ in (\ref{su5bes}), and $HW$ indicates the 
specification given previously (\ref{D6}) of the highest 
weight state of the $su(5)$ irrep $(1,1,1,1)$.
 
Thus the four $L$-operators may be represented by a set of four Dirac matrices 
$\Gamma_{\mu}$, where $\mu \in \{ 1,2,3,4 \}$ with $\chi$ represented by
$\Gamma_5=\Gamma_1 \Gamma_2 \Gamma_3 \Gamma_4$. While the $\Gamma_{\mu}$
have complicated looking actions, although with the correct anti-commutation
relations 
\be \label{D24}
\{ \Gamma_{\mu} \, , \,\Gamma_{\nu}  \}=2\delta_{\mu \nu} \quad , \e
\noindent the bosonic scalar $\chi$ is given by
\be \label{D25}
\chi =(2N_{B1}-1) \, (2N_{B2}-1) \mapsto I \otimes \sigma_3  
\e
To compare with the action of $K_3$, $K_5$ and chirality on the highest
weight states (\ref{C10}), we recall that from Sec. 3 that these 
actions are given simply by Pauli matrices. Also the eigenvalues of chirality 
distinguish between the two octets that comprise the Brink-Ramond 
representation of $su(3)$. While we have found a fairly natural analogue of 
this picture for $su(5)$, no nice representation of the $\Gamma$ matrices 
just discussed emerges.

\section{The odd case}

Again doing the simplest odd case of $su(2)$ with $r=3$ and $l=1$ with
sufficient care indicates the pattern that governs the general odd case
quite clearly. 

  Given $S_c=-\frac{1}{4} i\epsilon_{abc} \ga_a \ga_b$, we have
\be
\label{E1} 2A=\ga_1 -i\ga_2 \; , \; 2A^\dagger=\ga_1 +i\ga_2 \; , \;
N_A=A^\dagger \, A \; , \; -i\gamma_1\gamma_2 = 2(N_A-1)\quad,
\e 

\be
\label{E2} S_3=-\fract{1}{2}i \ga_1 \ga_2 =N_A-\fract{1}{2} \; , \; 
S_+=\ga_3 A^\dagger \; , \; S_-=A\ga_3 \quad . 
\e \noindent
It then follows algebraically that ${\bf S}^2=\fract{3}{4} I$, 
without raising the question as to
whether or not we use Pauli matrices $\sigma_1$ and $\sigma_2$ for
$\ga_1$ and $\ga_2$. 
The $K_3$ operator is now given by

\be
\label{su(2)k}
K_3 = -\fract{i}{12}\epsilon_{abc}\ga_a\ga_b\ga_c
\e

One way to get a complete Fock space description 
introduces $\phi$ such that $\phi^2=1 \; , \; \phi^\dagger 
=\phi \; , \; \{\phi \, , \, \ga_a \} =0$,
for $a \in \{ 1,2,3 \}$. Defining $B$ by

\be
\label{E3} 2B= \ga_3 -i\phi \quad , 
\e 

\noindent
so that $N_B=B^\dagger \, B$, $-i\gamma_3\phi=2(N_B-1)$, 
we find the representation 

\be
\label{E4} S_+=(B+B^\dagger)  A^\dagger \; , \; S_-=(S_+)^\dagger \quad . 
\e

\noindent 
Now in the $\C^2 \otimes \C^2$ Fock space of the $B$ and $A$ 
fermions, we build two equivalent $j=\frac{1}{2}$ irreps of $su(2)$ of
chiralities $\pm$, using the definition of chirality 

\be \label{E5} 
\ga_5\equiv -\ga_1 \ga_2 \ga_3 \phi 
= (-1)^{N_B+N_A}=(-1)^{tot. Fermi \#} \quad,
\e

\noindent 
Labelling the Fock states $|n_B,n_A\rangle$, we set

\be
\label{E6} |\pm ,j=\fract{1}{2} \, m=-\fract{1}{2} \rangle =|\tatop{0}{1} ,0
\rangle \quad .
\e 

\noindent 
Then,

\be
\label{E7} |\pm , j=m=\fract{1}{2} \rangle =S_+|\pm ,j=\fract{1}{2} \, 
m=-\fract{1}{2} \rangle= \pm | \tatop{1}{0} ,1\rangle \quad \e 

\noindent
with a typical and essential fermionic minus sign making its 
appearance. It is clear that we have two orthogonal $j=\frac{1}{2}$ 
irreps. It is easy too to compute the matrices of our Fock space 
representation ($\sigma_0={\bf 1}_2$)

\be
\label{E8} 2S_3 \mapsto \sigma_0 \otimes \sigma_3 \; , \; 2S_1 \mapsto 
\sigma_3 \otimes
\sigma_1 \; , \; 2S_2 \mapsto \sigma_3 \otimes \sigma_2 \quad , \e
\be\label{E9} \ga_1 \mapsto \sigma_2 \otimes \sigma_2 \; , \;
\ga_2 \mapsto -\sigma_2 \otimes \sigma_1 \; , \;
\ga_3 \mapsto \sigma_1 \otimes \sigma_0 \; , \; \quad , \e
\be\label{E10} \phi =\ga_4 \mapsto \sigma_2 \otimes \sigma_3 \; , \;
\ga_5 \mapsto  \sigma_3 \otimes \sigma_0 \; , \; 
2K_3\mapsto  \sigma_1 \otimes \sigma_3
\quad , 
\e 

\noindent
with all the expected properties. The action of $K_3$ 
(eq. (\ref{su(2)k})) on any state of our Fock space 
basis is to produce a state of 
opposite chirality. We note that all quantities that should 
anticommute with other fermionic quantities actually do so.

  To achieve all the features noted above, a certain 
price has had to be paid. In case it may not be thought worthwhile 
to pay it, especially in simple contexts, we offer the 
following argument. To reach a Fock space description of
${\cal D}$ for $su(2)$, we chose to introduce a dynamical 
variable $\phi$ of Majorana type not present in the original
formalism. This led us to use $4\times 4$ rather than 
$2\times 2$ gammas. Had we used Pauli matrices, $\ga_a=\sigma_a$,
the Kostant operator (\ref{su(2)k}) would have read 

\be
\label{E11} K_3=-\fract{1}{12}i\epsilon_{abc} \sigma_a \sigma_b \sigma_c
=\fract{1}{2}\, {\bf 1}_2 \quad , 
\e 

\noindent
losing sight of the dynamical role of $K_3$, and of its
fermionic nature as well. Using the 
irreducible two-dimensional picture is much 
the same as using only the chirality plus piece of the system, 
and forgetting about the negative chirality piece. This does 
represent ${\cal D}$ itself more or less satisfactorily, but fails, 
as our formalism does not, in the treatment of a composite system 
of two independent systems of type ${\cal D}$ ~\cite{MM}. 

Passing to the general odd case, we see that it is only the 
last gamma
$\gamma_r$, viewed as a lone Majorana fermion, 
that makes us do 
more than we have already done for the even case. And it is to be 
treated just as we described for ($\ga_3,\phi$) in the
$su(2)$ example. Thus we have $(r-l)/2$ fermions of $A$-type 
as well as $(l+1)/2$ of $B$-type, allowing for $(l+1)/2$ copies 
of the irrep $(1,\mathop{\dots}\limits^l,1)$. 
The latter irrep has dimension 
$2^{(r-l)/2}$ and uses $r=2s+1$ gammas that may
represented minimally by $2^s\times 2^s$ matrices. This 
itself would imply that ${\cal D}$ contains $2^p$ copies of 
$(1,\mathop{\dots}\limits^l,1)$ with $p=s-(r-l)/2=(l-1)/2$.
But we have in our Fock space realisation of ${\cal D}$ twice 
as many copies in virtue of the fact that we chose to adjoin 
to the original dynamical system one additional gamma, 
thereby doubling the size of all the gammas. We have the 
choice for ${\cal D}$ itself of accepting this, or else, 
of using only the positive chirality half of the full 
fermionic Fock space that carries $2^{(l-1)/2}$ copies of 
$(1,\mathop{\dots}\limits^l,1)$ for $\g$. As above, this 
fails to bring into clear focus the full fermionic nature of all 
the relevant fermionic quantities for the system ${\cal D}$.

\section{The chirality operator $\gamma_{r+1}$ 
and final remarks}

The previous discussion has exhibited the existence of $l$ 
fermionic anticommuting scalars $K_{(2m_i-1)}$
which are constructed from the Lie algebra cohomology cocycles of a Lie 
algebra $\g$ of even dimension $r$,
and the associated set of Dirac matrices. 
In the even case, to which the rest of this section mainly refers,
all these fermionic scalar operators change the chirality
of the states, since they anticommute with $\gamma_{r+1}$.
In the simplest case of $su(3)$, we have seen in eq. (\ref{C23A}) 
that $-iK_3 K_5 =\gamma_9$, while for $su(5)$ we see, 
from (\ref{D20})-(\ref{Dgamma25}), that for actions on highest 
weight states
\be 
\label{Dgam25}
K_3\,K_5\,K_7\,K_9\,\propto \chi = \gamma_{25} 
= - \prod_{\alpha=1}^{24} \gamma_{\alpha}
\quad . 
\e 
\noindent
But, since all $su(5)$ generators and in particular the 
raising and lowering operators commute with the ($\g$-invariant) 
$K$'s, the same applies to actions on all states. In general, 
we expect, for each even $\g$, that $\gamma_{r+1}$ is given to 
within normalisation by the product of all the available $K$'s, 
$\prod_{i=1}^{l}K_{(2m_i-1)}$; notice that 
$\sum_{i=1}^{l} (2m_i-1)=r$.

  The above properties of the various $K$'s reflect
the underlying group geometry. The $l$ cocycles of the 
Lie algebra cohomology of $\g$ may be looked at as invariant 
forms on the compact group manifold $G$ associated with $\g$ 
and, although not all the form properties are transported to
the $K$'s (for instance, unlike forms, $K^2\propto 1$), 
some of them are. The product of all the $(2m_i-1)$-forms associated 
with the cocycles is the volume $r$-form on $G$, an even form 
for each even $\g$. This accounts for duality 
relations such as (\ref{C23A}), which can be read 
as $K_3 (* K_3)\propto \gamma_9$, expressing the fact that
$K_3$ and $K_5$ are dual to each other. It also shows
that, when defined, the even chirality operator takes over 
the role of the volume form on the group manifold in 
the present context.

  The original motivation of ~\cite{brra} was to study
the invariant cubic Kostant operator on Lie algebra (symmetric) 
cosets, to understand the physical degrees of freedom
of certain supersymmetric theories; these appeared as
solutions to the Kostant-Dirac equation associated 
with specific cosets. The generalisations introduced in
this paper retain many of the properties of the 
representation independent part of the cubic Kostant 
cubic operator, in particular that their square is given by
Casimir invariants. At the same time, however, they have
rich geometrical properties that reflect their Lie algebra 
cohomology origin (as {\it e.g.}, that the product of the $l$
operators $K_i$ in the even case is represented by the
chirality/volume form). It seems worthwhile
to extend the geometrical methods of this paper
to the coset case, not discussed here, and to the possible 
full higher order Kostant operators. This, and the 
analysis of the hidden supersymmetries mentioned in
the introduction, will be discussed elsewhere.

\vskip 1.5cm

\noindent
{\bf Acknowledgements}. This work was partly supported by the
DGICYT, Spain ($\#$PB 96-0756) and PPARC, UK.

\end{document}